\documentstyle[amssymb,pra,aps,twocolumn,epsfig]{revtex}

\begin{document}
\title{Nondistortion Quantum Interrogation using EPR entangled Photons}
\author{Xingxiang Zhou$^{1,2}$, Zheng-Wei Zhou$^1$, Marc J. Feldman$^2$, Guang-Can
Guo$^1$\\
{\em $^1$Laboratory of Quantum Communication and Quantum Computation
and Department of Physics,\\
University of Science and Technology
of China, Hefei, Anhui 230026, China \\
$^2$Superconducting Electronics Lab, Electrical and Computer Engineering
Department, \\
University
of Rochester, Rochester, NY 14623, USA}}

\maketitle

\begin{abstract}
We propose a novel scheme for nondistortion quantum interrogation
(NQI), defined as an interaction-free measurement which preserves
the internal state of the object being detected. In our scheme,
two EPR entangled photons are used as the probe and polarization
sensitive measurements are performed at the four ports of the
Mach-Zehnder interferometer. In comparison with the previous
single photon scheme, it is shown that the two photon approach has
a higher probability of initial state preserving interrogation of
an atom prepared in a quantum superposition. In the case that the
presence of the atom is not successfully detected, the experiment
can be repeated since the initial state of the atom is
unperturbed.
\end{abstract}





One manifestation of the peculiar wave-particle duality of the
light quantum is the possibility of interaction-free measurements,
in which the presence of a classical or quantum mechanical object
in an interferometer path can be inferred without apparent
interaction with the probe photon. This is possible because the
presence of an object modifies the interference between different
branches of the photon wave function, so that there is a finite
probability that the photon will exit the interferometer through a
port where it should have not appeared in absence of the object.
This is the idea of interaction-free measurement (IFM) first
proposed by Elitzur and Vaidman \cite{ref:EV93}. Later Kwiat {\em
et al.} showed that the efficiency of IFM can be brought
arbitrarily close to 1 if one takes advantage of a discrete form
of the quantum Zeno effect \cite{ref:Kwiat95} \cite {ref:Kwiat99}.
More recently, Mitchison
and Massar 
proved that interaction-free discrimination between
semi-transparent and complete transparent (absent) objects can
also be done with probability approaching unity \cite{ref:semi}.

It is interesting to ask what effect does the interaction-free
measurement have on the object being detected, even though the
measurement is ``interaction free". As emphasized by Vaidman
\cite{ref:free?}, since the interaction Hamiltonian does not
vanish, in general the IFM can change very significantly the
quantum state of the observed object. As a matter of fact, if the
wave function of the observed object was initially spread out in
the space,
$|\psi\rangle=|\psi_{spatial}\rangle|\psi_{internal}\rangle$, then
a successful IFM necessarily collapses the spatial part of the
wave function to the vicinity of the optical path. However, in
most cases it is advantageous if we can keep the internal state of
the object unperturbed and realize an (internal) initial state
preserving IFM, which we may call a nondistortion quantum
interrogation (NQI) \cite{ref:note}. For a classical object or two
level atom in the ground state, a successful IFM is also a
nondistortion interrogation since the internal state of the object
is not affected if the probe photon is not absorbed. However, as
discussed in a recent paper by Potting {\em et al.}
\cite{ref:main}, this problem is more subtle for a quantum
mechanical object characterized by its quantum superposition.
Using a ``which path'' argument they showed that the absence of
energy exchange during the measurement is not a sufficient
condition to preserve the initial state of the object, since the
quantum superposition of the object is subject to measurement
dependent decoherence. Although it is possible in general to
design interaction-free measurement schemes that do preserve the
initial state of the quantum object, the scheme they discussed has
a very low success probability.
\begin{figure}[h]
    \centering
    \epsfig{file=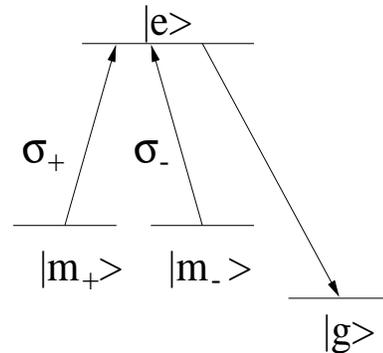, width=2in, height=1.8in}
\caption{Level structure of the atom. The atom can make a transition to the
excited state $|e\rangle$ from $|m_+\rangle$ or $|m_-\rangle$ by absorbing a
circular polarized photon. It then decays rapidly to the stable ground state
$|g\rangle$.}
\label{fig:atom}
\end{figure}

The purpose of this paper is to propose a new scheme for NQI,
which uses a pair of probe photons. Consider a multilevel atomic
system shown in Fig. \ref{fig:atom}, which is the same model as
used in \cite{ref:main}. Starting from the initial degenerate
metastable states $|m_+\rangle$ and $|m_-\rangle$, the atom can
make a transition to the excited state $|e\rangle$ by absorbing a
$+$ or $-$ (circular) polarized photon with unit efficiency. It
then decays irreversibly to the ground state $|g\rangle$ very
rapidly. The absorption process is therefore
\begin{equation}
\hat{a}^\dagger_\pm |0\rangle|m_\pm\rangle\longrightarrow |S\rangle|g\rangle
\end{equation}
where $|S\rangle$ is a scattered photon which we assume will not
be reabsorbed by the atom and can be filtered away from the
detectors. To investigate the effect of IFM on the initial state
of the atom, let us assume that the atom is initially in the
superposition
\begin{equation}
|\psi_{atom}\rangle=\alpha|m_+\rangle+\beta|m_-\rangle
\end{equation}
where $\alpha$ and $\beta$ are unknown non-vanishing coefficients
satisfying $|\alpha|^2+|\beta|^2=1$.

\begin{figure}[h]
    \centering
    \epsfig{file=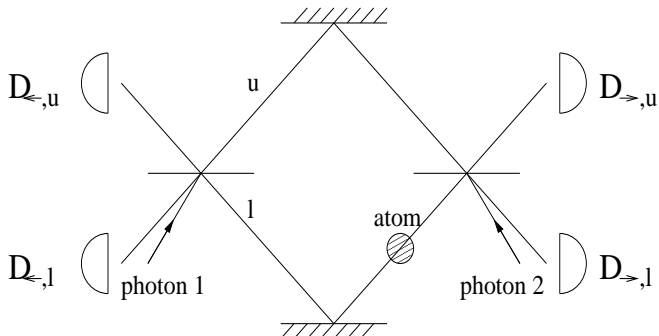, width=3.4in, height=1.75in}
\caption{Experimental setup of the two-photon nondistortion
quantum interrogation. The two probe photons enter the
Mach-Zehnder interferometer from left and right lower port
respectively. Polarization sensitive measurements of the photons
are performed at the four ports of the interferometer. }
\label{fig:interf}
\end{figure}

As illustrated in Fig. \ref{fig:interf}, the Mach-Zehnder
interferometer consists of two identical non-polarizing $50-50$
beam splitters. Note that the interferometer has four ports. We
use two photons, one entering from the left lower port and the
other from the right lower port, as our probe. Four polarization
sensitive photon detectors, $D_{\leftarrow,u}, D_{\leftarrow,l},
D_{\rightarrow,u}, D_{\rightarrow,l}$, are placed at the four
ports of the interferometer. When no atom is in the
interferometer, the two photons exit with certainty from the two
upper ports (therefore $D_{\leftarrow,u}$ and $D_{\rightarrow,u}$
fire). In presence of the atom, the interference is modified so
that one or both photons have a chance to exit from the lower
ports. So in the case that no absorption happened, any of the
following combinations indicates the presence of the atom in the
interferometer (hence a successful IFM): $D_{\leftarrow,u}$ and $
D_{\rightarrow,l}$ fire; $D_{\leftarrow,l}$ and
$D_{\rightarrow,u}$ fire; $D_{\leftarrow,l}$ and
$D_{\rightarrow,l}$ fire. These are not necessarily nondistortion
interrogation though. We will show, by properly choosing the
polarization of the probe photons and photon detectors, we can
perform a nondistortion interrogation of the atom with success
probability half that in the original EV scheme for a two level
atom \cite{ref:EV93}.

At first glance, it appears that the two photon approach is the
same with doing the measurement twice with the single photon
scheme as in \cite {ref:main}, once with a left entering photon
and once with a right entering one. This is not true for two
reasons. First, we can have correlations between the two probe
photons. Second, in the two photon scheme the states of the two
photons are measured together at the end of the experiment and the
wave function of the whole system (atom plus photons) is collapsed
only once. To make it clearer let us first look at the case that
two independent photons are used as the probe. Take the state of
the two photons to be
\begin{equation}
|\psi_{photon}\rangle=\hat{a}^\dagger_{\rightarrow,l,+}\hat{a}%
^\dagger_{\leftarrow,l,-}|0\rangle
\end{equation}
Note three indices are used to specify the state of a photon: the
propagation direction ($\rightarrow$ or $\leftarrow$); the optical
path it follows ($l$ower or $u$pper); and the direction of
polarization ($+$ or $-$). The effect of the beam splitters on the
photons is as follows:
\begin{eqnarray}
\hat{a}^\dagger_{\rightleftarrows,l}|0\rangle & \longrightarrow & \frac{1}{%
\sqrt{2}}(\hat{a}^\dagger_{\rightleftarrows,u}\pm i\hat{a}%
^\dagger_{\rightleftarrows,l})|0\rangle  \nonumber \\
\hat{a}^\dagger_{\rightleftarrows,u}|0\rangle & \longrightarrow & \frac{1}{%
\sqrt{2}}(\hat{a}^\dagger_{\rightleftarrows,l}\pm i\hat{a}%
^\dagger_{\rightleftarrows,u})|0\rangle
\end{eqnarray}
each time the photon is reflected its wave function picks up a phase shift
of $\pm \pi/2$ depending on the propagation direction of the photon.

Suppose that the possible location of the atom is in the lower arm
of the interferometer, then only photons following the lower path
interact with the atom. In addition, assume that the decaying from
the excited state to the ground state is so rapid that we do not
need to consider the stimulated emission of the atom when it is in
the excited state. Then the initial state
$|\psi_{photon}\rangle|\psi_{atom}\rangle$ evolves into the
following final state:
\begin{eqnarray}
|\psi_{final}\rangle&=&+\frac{1}{2}\hat{a}^\dagger_{\rightarrow,u,+} \hat{a}%
^\dagger_{\leftarrow,u,-}|0\rangle(\alpha|m_+\rangle+\beta|m_-\rangle)
\nonumber \\
& &-\frac{i}{2}\hat{a}^\dagger_{\rightarrow,l,+} \hat{a}^\dagger_{%
\leftarrow,u,-}|0\rangle\alpha|m_+\rangle  \nonumber \\
& &+\frac{i}{2}\hat{a}^\dagger_{\rightarrow,u,+} \hat{a}^\dagger_{%
\leftarrow,l,-}|0\rangle\beta|m_-\rangle  \nonumber \\
& &+\frac{1}{\sqrt{2}}|scattered\rangle
\end{eqnarray}
where $|scattered\rangle$ is the normalized state vector
corresponding to the situation that the atom absorbed one probe
photon and emitted one scattered photon by decaying to the ground
state afterward. If the probing photons were not absorbed there
are three possible outcomes: $D_{\rightarrow,u}$ and
$D_{\leftarrow,u}$ fire; $D_{\rightarrow,l}$ and
$D_{\leftarrow,u}$ fire; $D_{\rightarrow,u}$ and
$D_{\leftarrow,l}$ fire. In the case that the photons are detected
by the two upper detectors, the presence of the atom is not
discovered, and the state of the atom remains unchanged so we can
repeat the experiment. In contrast, if only one $+$ or $-$
polarized photon is used, as shown in \cite{ref:main} the state of
the atom is changed even if its existence is not successfully
detected. In this sense the two photon scheme is closer to the
original EV proposal. Here we see that the two photon scheme is
not identical to doing the experiment twice with the single photon
method, because the atom interacts with the larger Hilbert space
spanned by the two probe photons and the whole system is measured
only once. When one of the lower detectors fires, we have
discovered the atom successfully without it absorbing the photon,
but it is not a nondistortion interrogation because the initial
superposition of the atom is destroyed and the atom is left in
either $|m_+\rangle$ or $|m_-\rangle$, with probabilities
$|\alpha|^2/4 $ and $|\beta|^2/4$ respectively. What happened is
that the wave functions of the atom and photons got entangled when
the photons were propagating through the interferometer and
partially absorbed by the atom, so that a measurement of the
photons projects the atom to a state which is different from its
initial superposition.

Let us go one step further and make use of an EPR entangled photon pair as
our probe:
\begin{equation}
|\psi _{photon}\rangle =\frac 1{\sqrt{2}}(\hat{a}_{\rightarrow
,l,+}^{\dagger }\hat{a}_{\leftarrow ,l,-}^{\dagger }+\hat{a}_{\rightarrow
,l,-}^{\dagger }\hat{a}_{\leftarrow ,l,+}^{\dagger })|0\rangle
\end{equation}
By the same considerations we can show that the final state of the system is
\begin{eqnarray}
&&\frac 1{2\sqrt{2}}(\hat{a}_{\rightarrow ,u,+}^{\dagger }\hat{a}%
_{\leftarrow ,u,-}^{\dagger }+\hat{a}_{\rightarrow ,u,-}^{\dagger }\hat{a}%
_{\leftarrow ,u,+}^{\dagger })|0\rangle (\alpha |m_{+}\rangle +\beta
|m_{-}\rangle )  \nonumber \\
&&+\frac i{2\sqrt{2}}(\hat{a}_{\rightarrow ,u,-}^{\dagger }\hat{a}%
_{\leftarrow ,l,+}^{\dagger }|0\rangle \alpha |m_{+}\rangle +\hat{a}%
_{\rightarrow ,u,+}^{\dagger }\hat{a}_{\leftarrow ,l,-}^{\dagger }|0\rangle
\beta |m_{-}\rangle )  \nonumber \\
&&-\frac i{2\sqrt{2}}(\hat{a}_{\rightarrow ,l,+}^{\dagger }\hat{a}%
_{\leftarrow ,u,-}^{\dagger }|0\rangle \alpha |m_{+}\rangle +\hat{a}%
_{\rightarrow ,l,-}^{\dagger }\hat{a}_{\leftarrow ,u,+}^{\dagger }|0\rangle
\beta |m_{-}\rangle )  \nonumber \\
&&+\frac 1{\sqrt{2}}|scattered\rangle
\end{eqnarray}
If we use $x$ and $y$ polarization
\begin{eqnarray}
\hat{a}_x^{\dagger }|0\rangle  &=&\frac 1{\sqrt{2}}(\hat{a}_{-}^{\dagger }-%
\hat{a}_{+}^{\dagger })|0\rangle   \nonumber \\
\hat{a}_y^{\dagger }|0\rangle  &=&\frac i{\sqrt{2}}(\hat{a}_{-}^{\dagger }+%
\hat{a}_{+}^{\dagger })|0\rangle
\end{eqnarray}
the final state can be rewritten as
\begin{eqnarray}
-\frac{1}{2\sqrt{2}}(\hat{a}^\dagger_{\rightarrow,u,x}
\hat{a}^\dagger_{\leftarrow,u,x}+\hat{a}^\dagger_{\rightarrow,u,y}
\hat{a}^\dagger_{\leftarrow,u,y})|0\rangle(\alpha|m_+\rangle+\beta|m_-\rangle)
 \nonumber \\
-\frac{i}{4\sqrt{2}}(\hat{a}^\dagger_{\rightarrow,u,x}
\hat{a}^\dagger_{\leftarrow,l,x}+\hat{a}^\dagger_{\rightarrow,u,y}
\hat{a}^\dagger_{\leftarrow,l,y})|0\rangle
(\alpha|m_+\rangle+\beta|m_-\rangle) \nonumber \\
+\frac{1}{4\sqrt{2}}(\hat{a}^\dagger_{\rightarrow,u,x}
\hat{a}^\dagger_{\leftarrow,l,y}-\hat{a}^\dagger_{\rightarrow,u,y}
\hat{a}^\dagger_{\leftarrow,l,x})|0\rangle
(\alpha|m_+\rangle-\beta|m_-\rangle) \nonumber \\
+\frac{i}{4\sqrt{2}}(\hat{a}^\dagger_{\rightarrow,l,x}
\hat{a}^\dagger_{\leftarrow,u,x}+\hat{a}^\dagger_{\rightarrow,l,y}
\hat{a}^\dagger_{\leftarrow,u,y})|0\rangle
(\alpha|m_+\rangle+\beta|m_-\rangle) \nonumber \\
+\frac{1}{4\sqrt{2}}(\hat{a}^\dagger_{\rightarrow,l,x}
\hat{a}^\dagger_{\leftarrow,u,y}-\hat{a}^\dagger_{\rightarrow,l,y}
\hat{a}^\dagger_{\leftarrow,u,x})|0\rangle
(\alpha|m_+\rangle-\beta|m_-\rangle) \nonumber \\
+\frac{1}{\sqrt{2}}|scattered\rangle
\end{eqnarray}

Now we see that if the polarizations of the 4 detectors are chosen
to be $x$, $y$ instead of $+$ and $-$, there are 5 possible
outcomes if no photon was absorbed: $(a)$ the photons are detected
by $D_{\rightarrow,u}$ and $D_{\leftarrow,u}$ in the same
polarization ($x$ or $y$); $(b)$ the photons are detected by
$D_{\rightarrow,u}$ and $D_{\leftarrow,l}$ in the same
polarization; $(c)$ the photons are detected by
$D_{\rightarrow,u}$ and $D_{\leftarrow,l}$ in different
polarizations; $(d)$ the photons are detected by
$D_{\rightarrow,l}$ and $D_{\leftarrow,u}$ in the same
polarization; $(e)$ the photons are detected by
$D_{\rightarrow,l}$ and $D_{\leftarrow,u}$ in different
polarizations. Among them, in case $(b)$ and $(d)$ the atom is
left in its initial superposition and a successful NQI has been
realized. The probability of a successful NQI is 1/8, twice higher
than that of the single photon scheme as in \cite{ref:main}. In
case $(c)$ and $(e)$, the existence of the atom is also detected,
but there is a phase shift of $\pi$ in the superposition of the
atomic state. The probability for such an event is also $1/8$. If
the photons are received by the two upper detectors (with
probability 1/4), they must have the same polarization and the
initial superposition of the atom is unperturbed. In this case the
presence of the atom is not discovered and the experiment can be
repeated. Also, we note that $D_{\rightarrow,l}$ and
$D_{\leftarrow,l}$ never fire together. This is a consequence of
the polarization selective photon-atom interaction. When the
photons passed through the atom their wave functions got entangled
if no photon was absorbed by the atom. The interference between
the upper and lower branches of the photon wave function in the
interferometer is such that the two photons never both exit from
the lower ports.

In the above observation, we see that the correlation in the probe
system and the joint measurement of the states of the photons are
keys to the
nondistortion interrogation of the atom with an increased success
probability. By using an EPR entangled photon pair, we effectively
expanded the Hilbert space spanned by the probe system. The joint
measurement of the states of the two photons with linearly
polarized photon detectors allows us to map half of the
object-discovering results to initial superposition preserving
interrogations.

We should point out that in the single photon scheme as in
\cite{ref:main}, when the upper (polarization sensitive) detector
fires, it is still possible to find out the existence of the atom
if the polarization of the probing photon was changed. In that
case the initial atomic superposition was changed too. This is
possible because polarization of the photon provides an additional
degree of freedom. If a right going $x$ polarized photon is used
as the probe, the process is
\begin{eqnarray}
\hat{a}^\dagger_{\rightarrow,l,x}|0\rangle(\alpha|m_+\rangle+\beta|m_-%
\rangle)\longrightarrow  \nonumber \\
-\frac{3}{4}\hat{a}^\dagger_{\rightarrow,u,x}|0\rangle(\alpha|m_+\rangle+%
\beta|m_-\rangle)  \nonumber \\
+\frac{1}{4}i\hat{a}^\dagger_{\rightarrow,u,y}|0\rangle(\alpha|m_+\rangle-%
\beta|m_-\rangle)  \nonumber \\
+\frac{1}{4}i\hat{a}^\dagger_{\rightarrow,l,x}|0\rangle(\alpha|m_+\rangle+%
\beta|m_-\rangle)  \nonumber \\
-\frac{1}{4}\hat{a}^\dagger_{\rightarrow,l,y}|0\rangle(\alpha|m_+\rangle-%
\beta|m_-\rangle)  \nonumber \\
+\frac{1}{2}|scattered\rangle
\end{eqnarray}
%

To discuss this in a more general framework, we can think of NQI
as the discrimination between the absence and presence of quantum
coupling (interaction Hamiltonian) between the object and probe
system without destroying the probe particles. When the wave
functions of the object and probe overlap in space, the coupling
is present and the two parts which were initially independent (the
probe and object) will be entangled. By carefully designed unitary
operations and measurements on the probe system one tries to
obtain information on the presence of the coupling without
disturbing the internal state of the object. From this point of
view, a good NQI scheme is one which is most likely to keep the
internal state of the object unchanged while simultaneously bring
the probe system to a state orthogonal to that corresponding to no
coupling in some carefully chosen basis. What our result suggests
is that the probe system and the operations and measurements on it
have to be very carefully designed for the purpose of NQI, due to
the fragility of quantum superposition. Nevertheless, it can be
shown that NQI can be done for an object in quantum superposition
with efficiency approaching unity \cite{ref:future}. Whether a
more general consideration can be formulated and the connection to
quantum information is of further interest to us
\cite{ref:general} \cite{ref:Guo}.

In conclusion, we have provided a new scheme for nondistortion
quantum interrogation by using a pair of EPR entangled photons. It
enables us to monitor the presence of an object without destroying
its state of superposition. Due to the expanded Hilbert space of
the probe photons, it is shown that our scheme yields a higher
probability for successful nondistortion interrogation of the atom
than the single photon scheme. If the existence of the atom is not
successfully detected, the experiment can be repeated since the
atomic state is unperturbed. This method of nondistortion
interrogation of quantum objects characterized by its
superposition may find its application in future quantum
information processing.

Work of Z.W.Z. and G.C.G. was funded by National Natural Science Foundation
of China. Research of X.Z. and M.J.F. was supported in part by ARO grant
DAAG55-98-1-0367.


\begin{references}
\bibitem{ref:EV93}  A. C. Elitzur and L. Vaidman,\ Found. Phys. {\bf 23},
987 (1993).

\bibitem{ref:Kwiat95}  P. Kwiat, H. Weinfurter, T. Herzog, A. Zeilinger, and
M. A. Kasevich, \ Phys. Rev. Lett {\bf 74}, 4763 (1995).

\bibitem{ref:Kwiat99}  P. G. Kwiat, A. G. White, J. R. Mitchell, O. Nairz,
G. Weihs, H. Weinfurter, and A. Zeilinger,\ Phys. Rev. Lett. {\bf 83}, 4725
(1999).

\bibitem{ref:semi}  G. Mitchison, S. Massar, \pra {\bf 63}, 032105 (2001).

\bibitem{ref:free?}  L. Vaidman, \ quant-ph 0006077

\bibitem{ref:note}  The term ``quantum interrogation'' was first suggested
by Kwiat {\em et al.}\cite{ref:Kwiat99}. Here we are using the
word ``nondistortion'' to emphasize the internal state preserving
nature of the measurement. 

\bibitem{ref:main}  S. Potting, E. S. Lee, W. Schmitt, I. Rumyantsev, B.
Mohring and P. Meystre,\ Phys. Rev. A {\bf 62}, 060101(R), 2000.


\bibitem{ref:future}  X. Zhou {\em et al.}, to be published.

\bibitem{ref:general} Z-W. Zhou {\em et al.}, unpublished.

\bibitem{ref:Guo} G-C. Guo and B-S. Shi, Phys. Lett. A 256, 109,1999.
\end{references}
\end{document}